\documentclass[12pt,preprint]{aastex}

\shorttitle{Time Variation in NGC~7538~IRS1}
\shortauthors{Franco-Hern\'andez \& Rodr\'\i guez}

\begin{document}

\title{Time Variation in the Radio Flux Density from the
Bipolar Ultracompact H~II Region NGC~7538~IRS1}

\author{Ramiro Franco-Hern\'andez and Luis F. Rodr\'\i guez} 
\affil{Centro de Radioastronom\'\i a y Astrof\'\i sica, UNAM, 
Apdo. Postal 3-72 (Xangari), 58089 Morelia, Michoac\'an, M\'exico}
\email{r.franco, l.rodriguez@astrosmo.unam.mx}

\begin{abstract}

We present high angular resolution ($\sim0\rlap.{''}1$-$0\rlap.{''}4$)
VLA observations at 2 and 6 cm made in 1983, 1986, and 1995
toward the ultracompact bipolar H~II region NGC~7538~IRS1.
We find, at both wavelengths, clear evidence of a decrease in the emission
from the lobes. This decrease, of orden 20-30\%, has not been observed 
previously in any ultracompact H~II region. Most likely, it is due
to recombination of the ionized gas in the lobes as a result of
a decrease in the available ionizing photon flux. It is unclear
if this decrease in the ionizing photon flux
is due to an intrinsic change in the exciting star
or to increased absorption of ionizing photons in the optically-thick
core of the nebula.
\end{abstract}  


\section{Introduction}

The ultracompact H~II (UC HII) regions
are small (diameters $<$ 10$^{17}$ cm) and
dense (electron densities $>$ 10$^{4}$ cm$^{-3}$) structures of gas that are 
maintained ionized by deeply embedded, recently formed O stars.  
They have a few recurring morphological types, one of them is the
bipolar type that comprises only a handful of objects (Churchwell 2002).
The bipolar UC HII 
regions have an hourglass shape when projected
in the sky, the waist is believed to be produced by
confinement by a disk or torus of neutral gas, while the lobes
contain outflowing gas.
  
One of the most studied bipolar UC HII regions 
is NGC 7538~IRS1. It was first found in the infrared  by
Wynn-Williams, Becklin \& Neugebauer (1974), and  later its
free-free emission was clearly
resolved  at centimeter wavelengths   
by Campbell (1984). Its radio spectrum becomes fully optically thin
only above $\sim$100 GHz, with a total flux density of
$\sim$1.6 Jy at this frequency (Akabane et al. 1992). At a 
distance of 2.8 kpc (Sandell, Wright, \& Forster 2003), this 
flux density implies 
an ionizing photon rate of $1.5 \times 10^{48}$ s$^{-1}$, that could be provided
by an O8.5 ZAMS star with luminosity of $5 \times 10^{4}~L_\odot$ (Thompson 1984).

There is a large scale molecular outflow in the
region (CO: Campbell \& Thompson 1984; HCN$^+$: Batrla, Pratap \&
Snyder 1988), that was proposed to be associated with NGC~7538~IRS1, but
observations made by Batrla, Pratap \& Snyder (1988) suggest that the
source that is driving the outflow is located $\sim$15$''$ 
south of IRS1. Gaume et~al.  (1995) made H66$\alpha$ recombination 
line  observations  and found that this source has one of the most broad
profiles ($\Delta v \simeq$ 150 km s$^{-1}$) seen in UC HII regions. 
There are also several masers
associated with the source (e.g. OH: Hutawarakorn \& Cohen,
2003; CH$_3$OH: Minier, Booth \& Conway, 2000 ; H$_2$O: Kameya et~al. 
1990; $^{15}$NH$_3$: Gaume et~al. 1991), among them there is the rare
formaldehyde (H$_2$CO) 4.83 GHz maser (Hoffman et~al. 
2003). Observations in HCN, HCO$^+$ (Pratap, Batrla \& Snyder 1991) 
and CS (Kawabe et~al. 1992) show the presence of 
denser material around the source;
this can account for the confinement of the outflow seen in radio,
especially in the south direction.

In this paper we present the analysis of archival VLA
continuum
observations of NGC 7538~IRS1 made with high angular resolution.
The main goal of our study was to search for proper motions
or time variability in this compact object.
In $\S$2 we describe the observations; in $\S$3 we discuss them,
and finally in
$\S$4 our main conclusions are given.

\section{Observations}

The archive observations of the NGC~7538 IRS1 region 
were made using the VLA of the 
NRAO\footnote{NRAO is
a facility of the National Science Foundation
operated under cooperative
agreement by Associated Universities, Inc.}\ in the 
A configuration. The 2 cm observations were made in
four epochs, while those at 6 cm were made in
two epochs (see Table 1).
The observations were made in both circular
polarizations with an effective bandwidth of 100 MHz.  The data were
edited and calibrated following the standard VLA procedures and using the
software package AIPS.  After self-calibration we made cleaned images of the
region with the ROBUST parameter of IMAGR set to 0, to optimize
the compromise between angular resolution and sensitivity.  
The data at 2 cm was then cross-calibrated using
the techniques developed by Masson (1986).
After cross-calibration, we made images using as restoring beam
the average beam of the images made at different epochs (see Table 1).
These images were then subtracted to produce difference images
in order to search for small changes across time. 
We could not cross-calibrate the 6 cm data given the presence of bright, extended
sources in the primary beam. These extended sources are not well described by the data
(because it does not have short enough spacings), and since they
dominate the total emission of the region, cross-calibration at 6 cm was not
feasible.  

The cross-calibration method of Masson (1986) tries to minimize differences,
introduced by phase and amplitude errors, between
the two epochs being compared. This tends to equalize the total flux density
in the images being compared. However, 
even before cross-calibration, it was evident in
our data that the southern lobe of NGC 7538 IRS1 was
decreasing in flux density with time. We then introduced an additional criterion
in the analysis of the data. Since the central region of the source
is optically thick, we assumed that it did not changed significantly with
time and minimized its rms in the difference image, after the method
of Mart\'\i, Rodr\'\i guez, \& Reipurth (1998).
For this, after cross-calibration, we subtracted the 2 cm images
allowing one of them to have small shifts in position as well as a
small scaling in its absolute amplitude value. 
Then, the difference image, $\Delta I$, is given by

$$\Delta I = I_1(\alpha, \delta) - (1+ \epsilon) I_2(\alpha + \Delta \alpha,
\delta + \Delta \delta),$$

\noindent where $I_1$ and $I_2$ are the images for epochs
1 and 2, $\alpha$ and $\delta$ are the celestial coordinates,
$\Delta \alpha$ and $\Delta \delta$ the shifts introduced
(of order a few milliarcseconds), and $\epsilon$ is the scaling correction
(of order a few percent). In the case of the 6 cm images, cross-calibration
was not feasible and the images were treated using only the last criterion. 

\section{Discussion}

\subsection{The Bipolar UC HII NGC 7538 IRS1}

The 2 cm images of NGC~7538 IRS1 obtained for the
three epochs shown in Fig. 1 are very similar among them
and show the well-known
bipolar morphology first seen by Rots et al. (1981) and Campbell (1984).
The inner parts of the structure have a peak flux density 
of 26 mJy beam$^{-1}$ (average of the three epochs), that 
for a beam of $0\rlap.{''}13$, implies a brightness temperature of $\sim$8400 K.
Since we are observing free-free emission
from photoionized gas with an electron temperature
of this order, this result implies that these regions 
are optically thick.
Observations made at 1.3 cm by Gaume et al. (1995)
indicate clumpiness in this inner region and thus that the
brightness temperature is even higher in some positions.
We will assume an electron temperature of 10$^4$ K. 

The inner two peaks of the source are at $\sim 0\rlap.{''}1$ from the center.
At this position, we estimate the width of the structure to be also of
order $\sim 0\rlap.{''}1$. At a distance of 2.8 kpc, an angular size of
$0\rlap.{''}1$ corresponds to $1.4 \times 10^{-3}$ pc. Since in the inner
parts of NGC~7538~IRS1 the free-free opacity at 2 cm exceeds 1,
$\tau_{2~cm} \geq 1$, this implies an emission measure
$EM \geq 9.0 \times 10^8$ cm$^{-6}$ pc. Assuming a physical depth
of $1.4 \times 10^{-3}$ pc for the emission region, we derive a lower limit
for the electron density of $n_e \geq 8.0 \times 10^5$ cm$^{-3}$.
 
The lobes of the structure are markedly asymmetric, with the
northern one reaching peak flux density values of $\sim$6 mJy beam$^{-1}$,
while the southern one reaches peak flux density values 
of $\sim$15 mJy beam$^{-1}$.
Making similar assumptions that for the central regions, these
flux densities imply that the lobes reach moderate optical depths,
($\tau_{2~cm} \simeq$0.2-0.4), and electron densities in 
the order of $n_e \geq 4.0 \times 10^5$ cm$^{-3}$. 
Similar conclusions about the morphology of NGC~7538~IRS1 were 
obtained by Campbell (1984). 

The large electron densities of NGC~7538~IRS1 imply fast electron recombination times,
$t_e \simeq 1/(n_e~\alpha_B)$, where $\alpha_B = 2.6 \times 10^{-13}$ cm$^{-3}$ s$^{-1}$
is the recombination coefficient for $T_e = 10^4$ K.
We then obtain recombination times of order 0.15 years for the core and 0.3 years
for the lobes.

\subsection{Difference Images}

In Figure 1 we show the difference image for 1995.50-1983.89. 
The same features observed in this difference image are seen in
the difference images for 1995.50-1986.18 and for
1986.18-1983.89 (not shown here) but, as expected, with less intensity 
given the smaller time difference.

The most clear feature of the 1995.50-1983.89 difference image 
is the large negative region appearing at the position of the southern lobe.
Since the lobes have only moderate optical depths at 2 cm,
we interpret this negative region as a decrease in the free-free
emission. The decrease in flux density in the southern lobe
between the two epochs is
$\sim$30 mJy. This represents a large decrease, of $\sim$30\%,
in the flux density
of the southern lobe that went from about 98 mJy in
1983.89 to 68 mJy in 1995.50.
However, this large change in the lobe represents a relatively small
change in the total flux density of the source.
Since the total free-free flux density of NGC 7538 IRS1
at high frequencies is 1.6 Jy and assuming that it
is divided equally between the two lobes, we conclude that this decrease
represents a negative change of only $\sim$4\%.
Then, a relatively small change in the total ionizing photon flux of the
region, produces a large relative change in the lobes. This is due
to the fact that most of the ionizing photon flux is absorbed in the core
region, with only the final $\sim$10\% being absorbed by the lobes. 
Then, a small change of a few percent in the total ionizing flux
appears ``amplified'' in the lobes by a factor of $\sim$10.

There is also a negative region in the difference map associated
with the northern lobe, implying a decrease of $\sim$20\% between
1983.89 and 1995.50, consistent with what is
observed in the southern lobe. Despite our attempts to minimize the rms
of the central regions (the ``waist'') of the source, there are
two positive regions of emission left whose nature remains
uncertain.

To confirm this decrease in the emission of the radio lobes of
NGC 7538 IRS, we made difference maps in a similar way for the 
6 cm observations made in 1983.89 and 1995.65 (see Figure 2).
Again, negative regions associated with the
lobes appear in the difference map. This result is in a way surprising
because the lobes can reach free-free optical depths
in excess of 2 (and we should not be able to see through them). We
tentatively attribute the variation to clumpiness in the lobes
that could produce lines of sight with lower opacity.

Over approximately the same period of the continuum obsevations analyzed
here, Hoffman et al. (2003) report a systematic \sl increase \rm in the
emission of a bright 6 cm formaldehyde maser feature. It is unclear if
this systematic trend in the maser emission is related with that observed in the
continuum.

Finally, to check the reliability of our method we analyzed
2 cm images taken close in time, namely in 1995.50 and 1995.65.
As expected given the very short time interval
between observations, the difference image (Figure 3) 
does not show any significant residual structures (with the possible
exception of a marginal negative feature in the southern lobe).

\subsection{Possible Explanations for the Decrease in 
the Flux Density of the Lobes}

What could be causing the decrease in flux density in the lobes of
NGC 7538 IRS1? 
We start noting that, by several reasons,
the decrease cannot be caused by proper motions
of the ionized gas. First, proper motions of regions of emission 
appear in the difference image as a pair of features:
one negative
at the old position of the emitting region and
located closer to the central star, and a
positive one
at the present position of the emitting region
and located more distant from the central star 
(see, for example, Fig. 1 of Rodr\'\i guez et al. 2001). 
Furthermore, variations due to proper motions would imply 
displacements of order $0\rlap.{''}3$ (the width in the sky of the
southern lobe) on a time scale of a few years (the decrease
is evident even in the 1986.18-1983.89 difference image),
that would translate in velocities of order 1,800 km s$^{-1}$,
much larger than the velocities of $\sim$150 km s$^{-1}$
expected for the ionized outflow.
Finally, the morphology of the lobes remains similar between epochs
and the decrease in flux density appears more or less
simultaneously in an extended region.

The discussed characteristics of the flux density decrease suggest that,
more likely,
we are observing recombination of the ionized gas in the lobes due
to a decrease in the available ionizing photon flux.
As noted before, the recombination time of the ionized gas in
the lobes is only $\sim$0.3 years. We then favor as the explanation a
decrease in the ionizing photon flux that reaches the lobes.
This decrease could result from a decrease in the ionizing photon flux
produced by the central star. However,
from the evolutionary canonical models of
Bernasconi \& Maeder (1996), it can 
be shown that as a massive young star similar
to that required to ionize NGC 7538 IRS1 settles into the
main sequence, there are small changes in its ionizing photon
flux of order 10\%, but on timescales in the order of 5,000 years.
Then, it appears that these intrinsic changes in the
stellar ionizing photon flux cannot account for a
4\% decrease in only $\sim$11 years.   

Another possibility is that
the decrease observed in the lobes is due to increased absorption of
ionizing photons in the core of NGC 7538 IRS1.
For example, there could be increased injection
of gas from the neutral torus into the surroundings of
the central star, decreasing the amount of photons
available for the lobes. Unfortunately, we
cannot test this hypothesis directly given the large
free-free opacity of the core at centimeter
wavelengths. Future observations above 100 GHz (where
all the nebula is expected to be optically thin in the free-free) could
show that the \sl total \rm flux density of the region remains
constant, with the lobes decreasing and the core increasing correspondingly. 
The presence in the difference 2 cm
image of positive components in the core region
(see Fig. 1) suggests that this hypothesis could be correct.

\section{Conclusions}

Our main conclusions can be summarized as follows.

1) We analyzed data taken at 2 cm toward the bipolar UC HII region
NGC 7538 IRS1, finding that its lobes show a decrease in
flux density in the order of 20-30\% over a time interval of
11 years. The 6 cm data confirm this result.

2) This relatively large decrease in the 
emission from the lobes is tentatively interpreted as due
to recombination of the ionized gas, caused by a 
decrease in the available ionizing
photon flux.

3) At present it is unclear if this decrease in the ionizing photon
flux of the lobes is due to an intrinsic change in the exciting star or
to increased absorption of ionizing photons in the core of the object.


\acknowledgments

RFH and LFR acknowledge the support
of DGAPA, UNAM, and of CONACyT (M\'exico).
This research has made use of the SIMBAD database, 
operated at CDS, Strasbourg, France.

\clearpage

\begin{figure}
\epsscale{.80}
\plotone{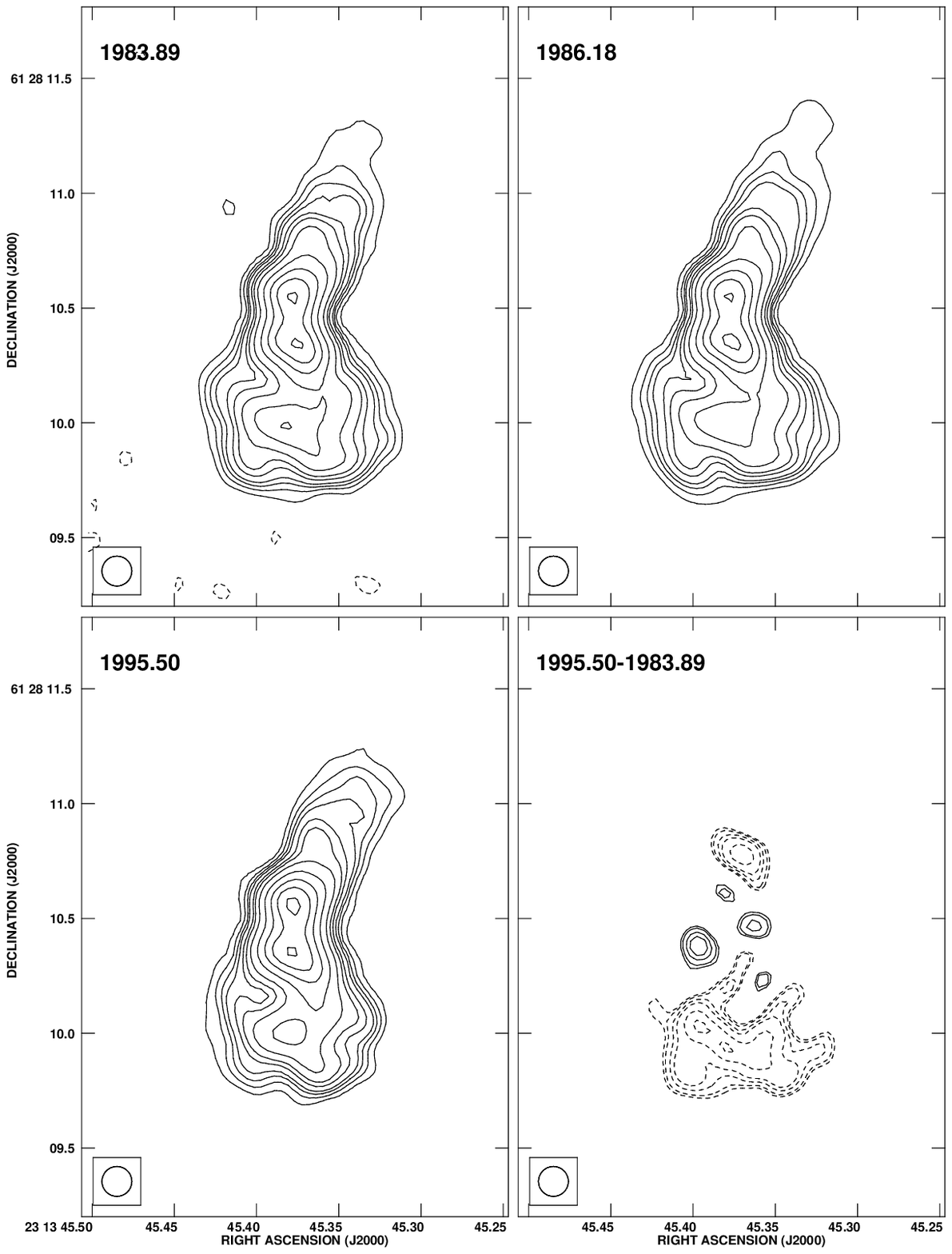}
\caption{Images of NGC 7538 IRS1 at 2 cm for epochs 1983.89, 1986.18,
and 1995.50, as well as the difference image for 1995.50-1983.89.
All images are reconstructed with a circular Gaussian beam
with HPBW of $0\rlap.{''}13$. The contours for the images of the individual
epochs are -5, 5, 10, 15, 20, 30, 40, 60, 100, 150, 200, and 250 
times 0.1 mJy beam$^{-1}$, the average rms of the three images.
The contours for the difference image are
-40, -30, -25, -20, -15, -10, -8, -6, -5, 5, 6, 8, 10, 15, and 20
times 0.2 mJy beam$^{-1}$.}
\end{figure}

\clearpage

\begin{figure}
\epsscale{.90}
\plotone{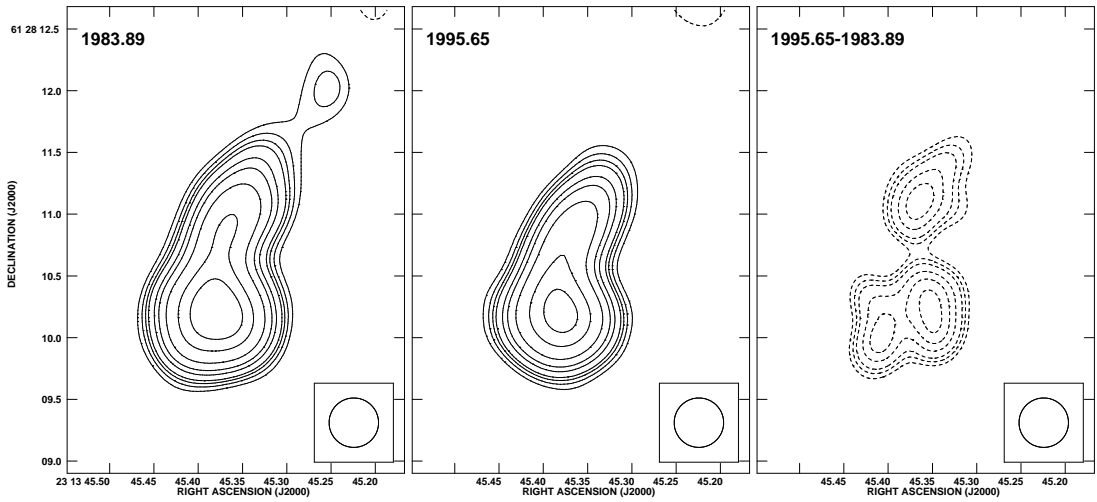}
\caption{Images of NGC 7538 IRS1 at 6 cm for the epochs 1983.89 
and 1995.65, as well as the difference image for 1995.65-1983.89.
All images are reconstructed with a circular Gaussian beam
with HPBW of $0\rlap.{''}4$. The contours for the images of the individual
epochs are 
-4, 4, 6, 8, 10, 15, 20, 30, 50, 70, and 100
times 0.23 mJy beam$^{-1}$, the average rms of the two images.
The contours for the difference image are
-20, -15, -12, -10, -8, -6, -5, -4, 4, and 5
times 0.27 mJy beam$^{-1}$.}
\end{figure}

\clearpage

\begin{figure}
\epsscale{.90}
\plotone{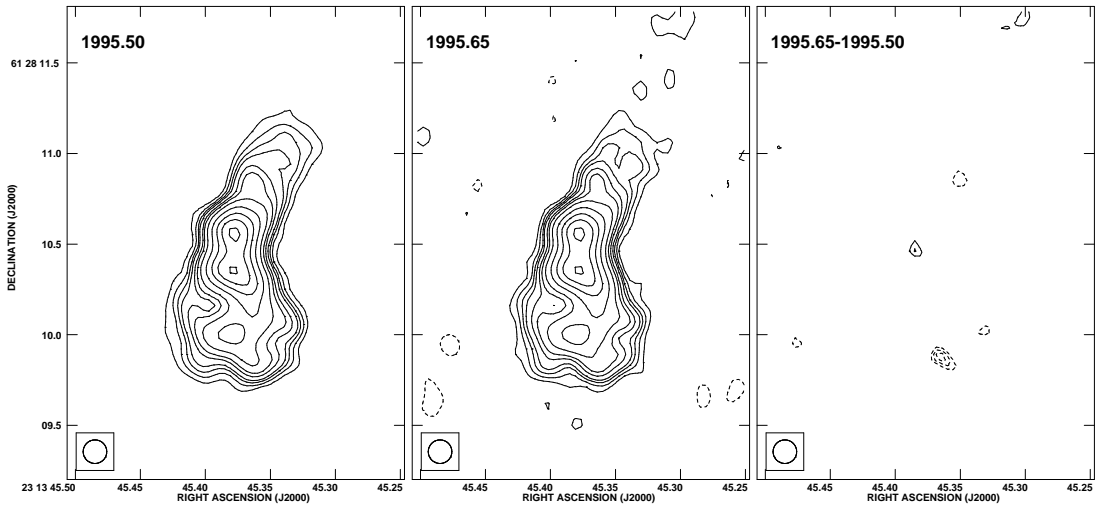}
\caption{Images of NGC 7538 IRS1 at 2 cm for the epochs 1995.50
and 1995.65, as well as the difference image for 1995.65-1995.50.
All images are reconstructed with a circular Gaussian beam
with HPBW of $0\rlap.{''}13$. The contours for the images of the individual
epochs are as in Figure 1.
The contours for the difference image are
-5, -4, -3, 3, 4, and 5
times 0.23 mJy beam$^{-1}$.}
\end{figure}

\clearpage

\begin{deluxetable}{ccccccc}
  \tabletypesize{\scriptsize}
  \tablecaption{Observational Parameters \label{tab:1} }
  \tablewidth{0pt}
  \tablehead{
 \colhead{Wavelength} &  & \colhead{Amplitude} & \colhead{Phase} &
 \colhead{Bootstrapped} & \colhead{Beam} & \colhead{rms noise} \\ 
  \colhead{(cm)}  & \colhead{Epoch} & \colhead{Calibrator}
 & \colhead{Calibrator} & \colhead{Flux Density (Jy)} & 
\colhead{($'' \times ''$;$^\circ$)} & \colhead{(mJy)}} 
\startdata
 2  & 1983 Nov 20 & 1331+305 & 2230+697  & 1.23$\pm$0.03 & 0.13$\times$0.12; +12 & 0.18 \\
 2  & 1986 Mar 06 & 1331+305 & 2230+697  & 0.94$\pm$0.01 & 0.15$\times$0.11; +85 & 0.07 \\
 2  & 1995 Jun 30 & 0137+331 & 2230+697  & 0.30$\pm$0.01 & 0.14$\times$0.12; $-$25 & 0.12 \\
 2 &  1995 Aug 26 & 0137+331 & 0019+734  & 0.92$\pm$0.02 & 0.13$\times$0.11; $-$6\phantom{0} & 0.21 \\[0.15cm]
 6  & 1983 Nov 20 & 1331+305 & 2230+697  & 1.43$\pm$0.01 & 0.41$\times$0.38; $-$33 & 0.22 \\
 6  & 1995 Aug 26 & 0137+331 & 0019+734  & 1.38$\pm$0.01 & 0.40$\times$0.37; +42 & 0.26 \\
\enddata
\end{deluxetable}

\end{document}